\documentclass[conference]{IEEEtran}

\usepackage{cite}

\ifCLASSINFOpdf
\else
\fi
\usepackage{array}
\usepackage{url}
\usepackage{graphicx}
\usepackage[tight,footnotesize]{subfigure}
\usepackage{listings, xcolor}
\usepackage{booktabs}

\begin{document}
%
\title{Toward Reverse Engineering of VBA Based Excel Spreadsheets Applications}

\author{\IEEEauthorblockN{Domenico Amalfitano, Nicola Amatucci, Vincenzo De Simone, Anna Rita Fasolino, Porfirio Tramontana}
\IEEEauthorblockA{Department of Electrical Engineering and Information Technologies\\ University of Naples Federico II \\ Via Claudio 21, Naples, Italy\\
Email: domenico.amafitano@unina.it, nicola.amatucci@unina.it, vincenzo.desimone2@unina.it, \\ anna.fasolino@unina.it, ptramont@unina.it}}


\maketitle

\begin{abstract}
Modern spreadsheet systems can be used to implement complex spreadsheet applications including data sheets, customized user forms and executable procedures written in a scripting language. These applications are often developed by practitioners that do not follow any software engineering practice and do not produce any design documentation. Thus, spreadsheet applications may be very difficult to be maintained or restructured. In this position paper we present in a nutshell two reverse engineering techniques and a tool that we are currently realizing for the abstraction of conceptual data models and business logic models supporting spreadsheet applications comprehension processes.

\end{abstract}
\IEEEpeerreviewmaketitle

\section{Introduction}
\label{introduction}
Spreadsheets are interactive software applications for manipulation and storage of data.  In a spreadsheet data are organized in worksheets, any of which is represented by a matrix of cells each containing either data or formulas that are automatically calculated at any variation of data. 
Modern spreadsheets systems (e.g. Microsoft Excel) are integrated with scripting languages (e.g., Microsoft Visual Basic for Applications). 
An end-user can extend the presentation layer of the application with Visual Basic for Application (VBA in the following) by defining new \textit{User Forms} and can develop new business functions by means of \textit{Procedures}. The execution of these procedures is event-driven: the procedures can be attached to events related to any element of the Excel Object Model, including Workbooks, Worksheets, Shapes and the same User Forms. Whenever one of these events occurs, the attached event handler code is executed. VBA can also be used to programmatically access and modify the underlying Excel Object Model, i.e. the hierarchy of objects contained in Excel representing all its accessible resources. 

Very often developers create complex spreadsheet applications without any documentation at design level, making hard any task related to their maintenance. In order to reconstruct design models of an existing spreadsheet application reverse engineering techniques and tools are needed.

The research community has devoted great attention to the analysis and comprehension of spreadsheet applications. However, most papers focus on analyzing and testing the formulas in a spreadsheet, while they leave out the analysis of the spreadsheet’s embedded code and of the static and dynamic relationships between the code and the spreadsheets cells.

In this paper we will present in a nutshell two reverse engineering techniques and a tool supporting the second technique allowing the abstraction of design models of an existing spreadsheet application. In details, in section \ref{dmre} we will present a technique for abstracting conceptual data models by analyzing spreadsheets data while in section \ref{smre} we will present a tool supporting the abstraction of views representing the relationships between VBA procedures, user forms and spreadsheet cells. Finally, in section \ref{conclusions} some conclusions and future works will be presented.

\section{Data Model Reverse Engineering}
\label{dmre}
The first technique presented in this paper regards the abstraction of conceptual data models from the analysis of the structure and of the information included in an Excel spreadsheet by means of heuristic rules. 

This technique is based on heuristic rules automatically applicable on a spreadsheet. By means of these rules, set of candidate classes with attributes, relationships between them and the corresponding cardinalities are abstracted on the basis of the structure and of the properties of spreadsheets and of their components, such as sheets, cells, cell headers, etc.. In particular, the heuristics consider cells labels and data by looking for repeated data, synonyms and group of cells containing well-defined data structures such as array strings, integer matrixes, etc. 

The considered rules has been presented in \cite{Amalfitano2014a} and \cite{Amalfitano2014} where they have been used with success to abstract the conceptual data model underlying some complex spreadsheet applications used in the automotive context. Some of the proposed rules were derived from the literature \cite{Abraham2006},\cite{Hermans2010}.

\begin{figure}[ht]
\centering
\includegraphics[width=\columnwidth]{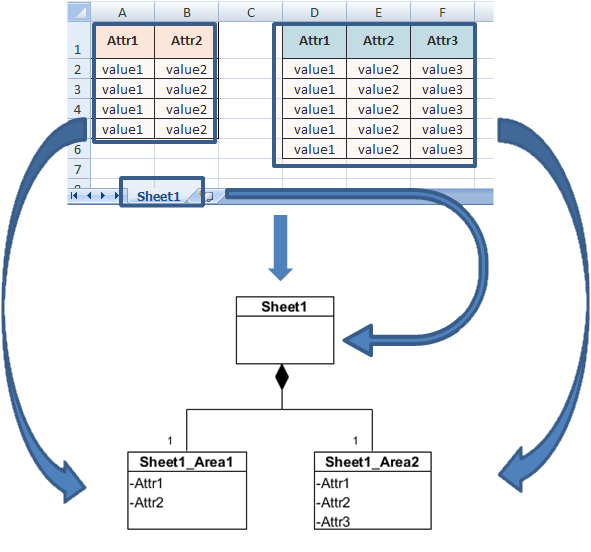}
\caption{An example of application of heuristic rules to abstract a conceptual data model of a spreadsheet}
\label{fig:dmre}
\end{figure}

Figure \ref{fig:dmre} shows an example of a possible application of some of the proposed rules. First of all, the sheet of the spreadsheet shown in Figure \ref{fig:dmre} may be abstracted as a class. Moreover, the two distinct rectangular areas separated by a blank column and composed of labels and data may be abstracted as two classes. Two composition relationships between these two classes and the class representing the sheet may be abstracted, too. The first rows of the two areas have cells with bold texts and different background colors: they may be abstracted as attributes of the classes representing the areas. The texts of these cells may be abstracted as names of the attributes of the corresponding classes. 

\section{Business Logic Model Reverse Engineering}
\label{smre}
The second contribution presented in this position paper is related to a reverse engineering technique that we have realized to abstract models of the business logic of a spreadsheet application by statically analyzing sheets, user forms, VBA procedures and their inter-relationships. The technique is completely supported by an interactive Excel add-in called EXACT (EXcel Application Comprehension Tool) that we have realized and that is available at \url{https://github.com/reverse-unina/EXACT}. It provides features for the extraction of information and the abstraction of several views of an existing spreadsheet application.

The tool extracts information from a spreadsheet by exploiting the features offered by the Office Primary Interop Assemblies for the Excel Application that exposed the Microsoft Excel Object Library and by statically analyzing the source VBA code. The tool is able to reconstruct the set of elements composing the spreadsheet including sheets, user forms, event handlers, classes and procedures. Furthermore the tool recovers information related to the relationships between these elements such as procedure calls, relationships between events and the corresponding handlers, and dependencies between procedures and cells. 

\begin{figure}[ht]
\centering
\includegraphics[width=\columnwidth]{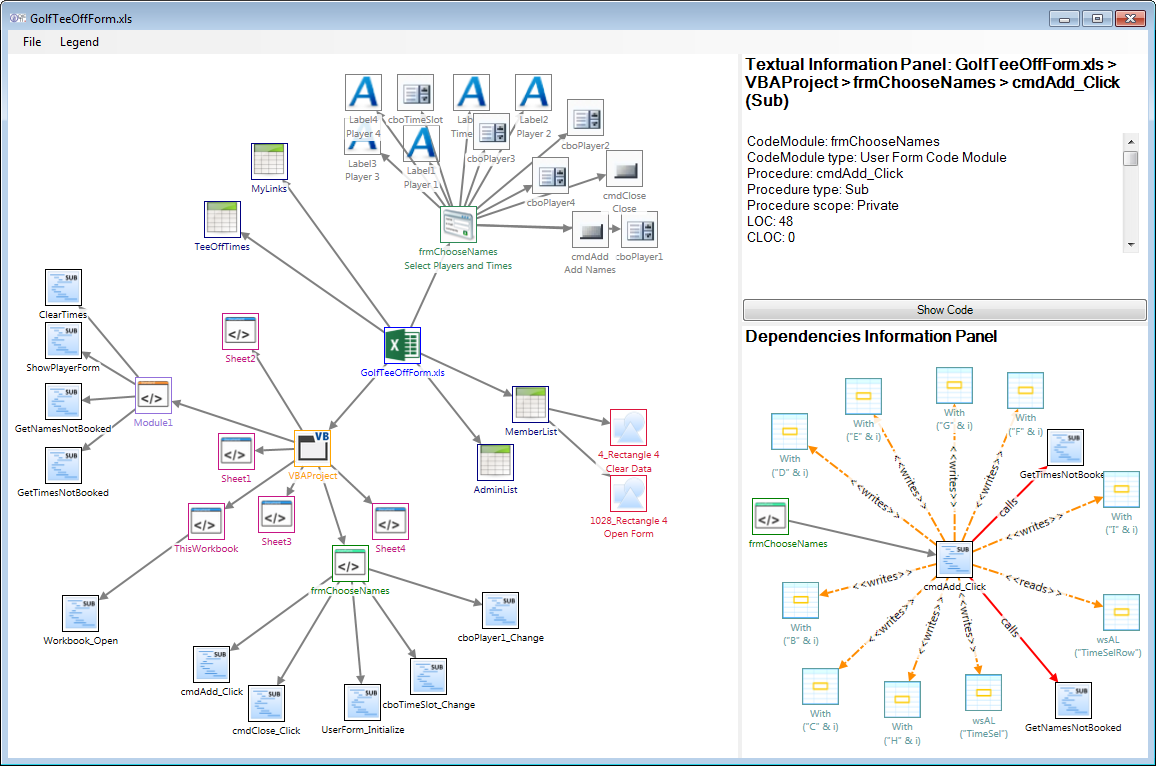}
\caption{Structural View of an Excel Spreadsheet}
\label{fig:smre}
\end{figure}

The EXACT tool provides several features of software visualization. It offers multiple views at different levels of detail and provides cross-referencing functions for switching between these views. In example, the left part of Figure \ref{fig:smre} exemplifies the structural view proposed by EXACT for an Excel spreadsheet. The figure shows that the spreadsheet is composed of a single workbook with 4 worksheets, a VBProject including 6 code modules (that implement a total of 10 procedures) and a user form with 12 controls. In the right part of the figure there are some detailed information provided by the EXACT tool about a procedure selected by the user and a graph showing the dependencies between this procedure and the other application components. In particular, the considered procedure writes in 9 different groups of cells, reads the values of a group of cells and calls two procedures.

\section{Conclusions and Future Works}
\label{conclusions}
In this position paper we have presented in a very concise way two reverse engineering techniques and a tool abstracting conceptual data models and business logic models of Excel spreadsheets, taking into account both the data structure, the VBA code procedures, the User Forms and their inter-relationships.
In future, we plan to extend the EXACT tool by implementing the data model reverse engineering technique proposed in section \ref{dmre} with the aim to make possible comprehension processes of Excel spreadsheets applications.

\bibliographystyle{IEEEtran}
\bibliography{sems2015} 

\end{document}